\documentclass[twocolumn,showpacs,preprintnumbers,amsmath,amssymb,pre]{revtex4}
\usepackage{graphicx}
\usepackage{dcolumn}
\usepackage{bm}

\begin{document}

\title{Frequency-dependent fluctuation-dissipation relations in granular gases}

\author{Guy Bunin$^1$}
\author{Yair Shokef$^2$}
\author{Dov Levine$^1$}
\affiliation{$^1$Department of Physics, Technion, Haifa 32000, Israel\\
$^2$Department of Physics and Astronomy, University of Pennsylvania, Philadelphia, PA 19104, USA}

\date{\today}

\begin{abstract}
The Green-Kubo relation for two models of granular gases is discussed. In the Maxwell model in any dimension, the effective temperature obtained from the Green-Kubo relation is shown to be frequency independent, and equal to the average kinetic energy, known as the granular temperature. In the second model analyzed, a mean-field granular gas, the collision rate of a particle is taken to be proportional to its velocity. The Green-Kubo relation in the high frequency limit is calculated for this model, and the effective temperature in this limit is shown to be equal to the granular temperature. This result, taken together with previous results, showing a difference between the effective temperature at zero frequency (the Einstein relation) and the granular temperature, shows that the Green-Kubo relation for granular gases is violated. 
\end{abstract}                            
                            

\maketitle

\section{Introduction}

Fluctuation-dissipation\ (FD) relations in non-equilibrium systems have been the subject of considerable attention recently \cite{cugliandolo,barrat2000a,perez_2003,bellon_2002,blickle_2007,harada2003,hayashi2005,chen}. In this context, granular matter serves as a perfect example of a non-equilibrium system. Such systems consist of macroscopic grains undergoing inelastic collisions. In a \emph{cooling state}, no energy is injected into the system, and the total kinetic energy decreases steadily, approaching a trivial steady-state where all grains are at rest. If energy is constantly supplied to the system a \emph{non-equilibrium steady-state} (NESS) is reached. This paper focuses on the frequency-dependent FD relations in this NESS.

If the density of grains is low, the system is known as a granular gas \cite{Poschel,poschel2003}. In this case, spatial correlations are small, and mean-field models, in which particles collide without regard to their positions, may provide good approximations of the system at hand. In such models, interactions  are governed by a stochastic collision rule, with some interaction rate. Two such models are studied in this paper, differing in their collision rules. The simpler of the two is the \emph{inelastic Maxwell model} \cite{Ernst,ben-naim}, in which the collision rate is taken to be constant for each pair of particles in the system. In the second, more realistic model, which we refer to as the \emph{mean-field granular gas }(MFGG), the collision rate of each pair of particles is proportional to the particles' relative velocity.

The FD theorem \cite{callen_welton,Kubo} connects the linear response of an equilibrium system to its correlations. Stated in the frequency ($\omega$) domain, it relates the response function to a weak harmonic force to the Fourier-Laplace transform of the auto-correlation function of the observable conjugate to that force. The theorem states that, remarkably, the two functions have \emph{exactly} the same frequency-dependence, and moreover, that their ratio is equal to the temperature of the system. Since the full frequency-dependent functions are not always easy to compute, their $\omega\rightarrow0$ limit is often calculated; the connection between fluctuation and response in this limit is known as the Einstein relation.

In a NESS, the response to a weak perturbation from the steady-state and the corresponding auto-correlation may be similarly defined. Since granular gases are out of equilibrium, however, the FD relation is not expected to generally hold, and the two functions may have different frequency dependencies. If we nevertheless define a FD temperature $T_{FD}$ as the ratio of the correlation and response functions \cite{Hohenberg89,cugliandolo}, we would expect this quantity to depend on frequency.

This having been said, there is evidence that the FD relation may nonetheless closely hold for driven granular gases. Molecular dynamics simulations of 2D granular gases~\cite{Puglisi2002} indicate that $T_{FD}$ is approximately frequency-independent, and moreover, is close to the average kinetic energy per particle, known as the granular temperature $T_{G}$. Einstein relations were calculated for the MFGG model, yielding a value for $T_{FD}$ different from $T_{G}$ \cite{garzo}. Numerically, however, this deviation was small, and consistent with direct Monte-Carlo simulations \cite{Barrat2004} that observed no violation of the FD relation (dependence of $T_{FD}$ on $\omega$) to within their accuracy. Experiments on air fluidized particles have shown that at low densities the effective temperature measured from the Einstein relations is close to the granular temperature \cite{abate}. For the 1D Maxwell model with a Gaussian thermostat, the Einstein relation was calculated \cite{baldassarri_2005} and yielded $T_{FD}=T_{G}$. Additional transport coefficients, calculated in \cite{Garzo_Astillero}, verify the Einstein relation in the Maxwell model. Subsequently, we showed that the full frequency-dependent FD relation holds exactly for the Maxwell model, driven by either a Gaussian or a stochastic thermostat \cite{shokef2006}. This result was recently shown to also hold in 2D with a Gaussian thermostat \cite{puglisi2007}. Fluctuation-dissipation relations in denser systems, in which correlations play a role, are discussed  e.g. in \cite{maske,barrat2002,ohern,danna_2003,puglisi2007}. Time dependent FD relations in other systems are studied in \cite{harada2003,hayashi2005,chen,fielding_sollich}.

In \cite{shokef2006} we also discussed a simple model of a granular gas \cite{srebro_levine}, for which the FD relations were calculated analytically. It was shown that in measurements where correlations between different degrees of freedom appear, the FD relation is violated. In \cite{puglisi2007} the role of correlations was studied in simulations of granular fluids, and was shown to affect the violation of the FD relation. However, FD violation has not yet been observed in simulations of dilute granular gases, due to their limited accuracy. In this paper we provide theoretical reasoning for why FD relations should be only slightly violated even in the dilute limit. 

We first show that the frequency dependent FD relation holds in the Maxwell model in any dimension and with both customarily used thermostats. We then show that the frequency-dependent FD relation is violated in the MFGG, where collision rates depend on velocity. Quantitatively, the violation is expected to be small, in accordance with previous simulation results.

\section{Fluctuation-dissipation relation}

The Green-Kubo (GK) formulation~\cite{Kubo} employs the frequency-dependent autocorrelation function, defined as the Fourier-Laplace transform of the velocity autocorrelation function,
\begin{equation}
D(\omega) \equiv \frac{1}{d} \int_{0}^{\infty} \langle \mathbf{v}(0) \cdot \mathbf{v}(t) \rangle e^{-i \omega t} dt, \label{eq:D_definition}
\end{equation}
where $d$ is the system dimension. Vectorial quantities are written in boldface font (e.g. $\mathbf{v}$ for the velocity). Note that $D(\omega)$ reduces to the diffusion coefficient in the zero frequency limit.

The frequency dependent mobility is calculated by applying a weak periodic force $\bm{\xi}(t)=\bm{\xi}_{0}e^{i\omega t}$ to a single particle $i$ and measuring the amplitude $\mathbf{A}\left(\omega\right)$ of its resulting velocity, averaged over the steady-state: $\left\langle\mathbf{v}_{i}\right\rangle =\mathbf{A}(\omega)e^{i\omega t}$. Then $\bm{\xi}_{0}$ and $\mathbf{A}$ are directed along the same direction, and the mobility $\mu(\omega)$ is defined so that
\begin{equation}
\bm{\xi}_{0}\mu(\omega)=\mathbf{A}\left(\omega\right).\label{eq:mu_definition}
\end{equation}
Equivalently, if some general time-dependent force $\bm{\xi}(t)$ is applied to particle $i$, and $\left\langle\mathbf{v}_{i}\left(t\right)\right\rangle$ is its resulting average velocity, $\mu\left(\omega\right)$ may be defined so that
\begin{equation}
\widetilde{\bm{\xi}}\left(\omega\right)\mu(\omega)=\left\langle\mathbf{\tilde{v}}_{i}\left(\omega\right)\right\rangle ,
\end{equation}
where $\widetilde{\bm{\xi}}\left(\omega\right)$ and $\left\langle\mathbf{\tilde{v}}_{i}\left(\omega\right)\right\rangle$ are the Fourier transforms of $\bm{\xi}(t)$ and $\left\langle\mathbf{v}_{i}\left(t\right)\right\rangle$ respectively. Note that both $D(\omega)$ and $\mu(\omega)$ are complex. 

In thermodynamic equilibrium the FD theorem, or Kubo formula, guaranties that $D(\omega)=T\mu(\omega)$, where the temperature is measured in units of energy, and Boltzmann's constant is set to one. The Einstein relation is obtained by taking the $\omega\rightarrow0$ limit, thereby connecting the mobility with respect to a constant force and the long time diffusion coefficient.

For a NESS one may use the fluctuation $D(\omega)$ and the response $\mu(\omega)$ to define an effective FD temperature \cite{cugliandolo,barrat2000,barrat2001,Berthier,garzo_dufty,danna_2003,ohern,Hohenberg89,shokef_levine_2006,srebro_levine}:
\begin{equation}
T_{FD}\left(\omega\right)\equiv\frac{D(\omega)}{\mu(\omega)}.\label{eq:T_FD_definition}
\end{equation}
$T_{FD}(\omega)$ scales fluctuations in the system and in general depends on the measurement frequency $\omega$. In this context, the FD theorem states that in equilibrium $T_{FD}\left(  \omega\right)  $ is independent of $\omega$ and is equal to the system's temperature.

\section{Inelastic Maxwell model}

In this section we show that in the Maxwell model, where the collision rate $\Gamma$ is velocity independent, the frequency dependent GK relations hold exactly with $T_{FD}=T_{G}$, where $T_{G}$, known as the \textquotedblleft granular temperature\textquotedblright, is defined as $T_{G}\equiv\left\langle\mathbf{v}^{2}\right\rangle/d$ (taking particles of unit mass). In~\cite{baldassarri_2005} the Einstein relations for the 1D Maxwell model with a Gaussian thermostat were shown to give $T_{FD}\left(0\right)=T_{G}$ (the thermostats are defined below). In \cite{shokef2006} we showed that for both Gaussian and  stochastic thermostats, in the 1D Maxwell model $T_{FD}\left(\omega\right)=T_{G}$, independent of frequency. For clarity of presentation, before turning to higher dimensions and to the more realistic MFGG model, we first present the derivation of this result in more detail.

In the 1D inelastic Maxwell model, each particle is described by its velocity $v_{i}$. Particles collide at a constant collision rate $\Gamma$, and the colliding particles are chosen at random. In a collision between particles $i$ and $j$, the velocity of each particle in the center of mass frame is multiplied by $-\alpha$, where $\alpha$ is the restitution coefficient. The particles are coupled to a thermostat, which can be of two types. A stochastic thermostat \cite{williams_mackintosh,van_noije} exerts a fluctuating force $\psi_{i}(t)$ and a velocity-dependent damping force $-\lambda v_{i}$ on each particle (with $\lambda>0$). During a finite time $\Delta t>0$, short compared to $\Gamma^{-1}$ and to $\lambda^{-1}$, the effect of the fluctuating force on a particle's velocity is proportional to $\sqrt{\Delta t}$ (see e.g.~\cite{williams_mackintosh}). In a Gaussian thermostat \cite{gauss_term}, instead of applying a random force $\psi_{i}(t)$, $\lambda$ is assigned a negative value such that the term $-\lambda v_{i}$ acts as a velocity-dependent driving force, which rescales the velocities of all particles in every time interval. The NESS achieved with a Gaussian thermostat is exactly the freely cooling state, rescaled at every time step \cite{montanero}. 

For both types of thermostats, the velocity of particle $i$ evolves according to
\begin{equation}
v_{i}(t+\Delta t)=\left\{
\begin{array}[c]{cc}
\text{\underline{value:}} & \text{\underline{probability:}}\\
(1-\lambda\Delta t)v_{i}+\psi_{i}\sqrt{\Delta t} & 1-\Gamma\Delta t\\
\frac{1-\alpha}{2}v_{i}+\frac{1+\alpha}{2}v_{j} & \Gamma\Delta t
\end{array}
\right.  ,\label{eq:dyn_rule_mm_v}
\end{equation}
where $v_{i},v_{j}$, and $\psi_{i}$ in the right-hand side are taken at time $t$. This represents the stochastic evolution of the system, where at probability $\Gamma$ per unit time, particle $i$ collides with a randomly chosen particle $j$ and their velocities are updated accordingly [second line in Eq. (\ref{eq:dyn_rule_mm_v})]. Keeping terms up to linear order in the short time interval $\Delta t$, the velocity of particle $i$ is updated due to the interaction with the thermostat only for time intervals during which it did not collide with another particle in the system [first line in Eq. (\ref{eq:dyn_rule_mm_v})]. For a stochastic thermostat, $\psi_{i}(t)$ is an uncorrelated random acceleration with $\langle\psi\rangle=0$, and $\langle\psi^{2}\rangle=2\lambda T_{B}$, where $T_{B}$ is a parameter, interpreted as the temperature of a bath to which the system is connected. For a steady-state of elastic particles ($\alpha=1$), $T_{G}\equiv\left\langle v^{2}\right\rangle =T_{B}$, as expected in equilibrium. For a Gaussian thermostat, $\psi_{i}(t)=0$ and $\lambda=(\alpha^{2}-1)\Gamma/4$, which maintains a constant $T_{G}$.

In order to obtain the velocity autocorrelation function, we multiply Eq.~(\ref{eq:dyn_rule_mm_v}) by $v_{i}(0)$ and average over stochasticity and initial conditions, to get
\begin{align}
& \left\langle v_{i}(0)v_{i}(t+\Delta t)\right\rangle = \nonumber\\ &  
\left[\left(1-\lambda\Delta t\right)\left\langle v_{i}\left(0\right) v_{i}\left(t\right)\right\rangle+\left\langle v_{i}\left(0\right)\psi_{i}\left(t\right)\right\rangle\sqrt{\Delta t}\right]\left(1-\Gamma\Delta t\right) \nonumber\\
& +\left[\frac{1-\alpha}{2}\left\langle v_{i}\left(0\right)v_{i}\left(t\right)\right\rangle +\frac{1+\alpha}{2}\left\langle v_{i}\left(0\right)v_{j}\left(t\right)\right\rangle\right]\Gamma\Delta t.
\end{align}
Noting that $\left\langle v_{i}(0)\psi_{i}(t)\right\rangle =0$, $\left\langle v_{i}\left(0\right)v_{j}(t)\right\rangle =0$, neglecting terms of order $\left(\Delta t\right)^{2}$, and rearranging, we take the limit $\Delta t\rightarrow0$ to obtain
\begin{equation}
\frac{d\langle v(0)v(t)\rangle}{dt}=-\kappa_{1D}\langle v(0)v(t)\rangle ,\label{eq:1d_mm_corr}
\end{equation}
where $\kappa_{1D}\equiv\lambda+\frac{1}{2}(1+\alpha)\Gamma$ is the \textquotedblleft effective drag\textquotedblright\ experienced by a particle due to the drag of the thermostat and the collisions with other particles. This differential equation has the solution
\begin{equation}
\langle v(0)v(t)\rangle=\langle v^{2}\rangle e^{-\kappa_{1D}t} .\label{eq:mox_mod_autocor}
\end{equation}
Using Eq. (\ref{eq:mox_mod_autocor}) in the definition of $D(\omega)$, Eq.~(\ref{eq:D_definition}), we find
\begin{equation}
D(\omega)=\frac{\langle v^{2}\rangle}{\kappa_{1D}+i\omega} .\label{eq:MM_D_calc}
\end{equation}

To calculate the mobility, a periodic acceleration $\xi(t)=\xi_{0}e^{i\omega t}$ is applied to a single particle $i$. Due to the mean-field assumption, and in the thermodynamic limit of a very large system, the probability distribution of $v_{j}(t)$ ($i \ne j$) is unaffected by the force $\xi(t)$ acting on particle $i$. From Eq.~(\ref{eq:dyn_rule_mm_v}), we similarly have
\begin{equation}
\frac{d\left\langle v_{i}(t)\right\rangle }{dt}=-\kappa_{1D}\left\langle v_{i}(t)\right\rangle +\xi_{0}e^{i\omega t}.\label{eq:langevin}
\end{equation}
This has the steady solution $\left\langle v_{i}(t)\right\rangle =\xi_{0}e^{i\omega t}/(\kappa+i\omega)$, so from Eq. (\ref{eq:mu_definition}) we see that
\begin{equation}
\mu(\omega)=\frac{1}{\kappa_{1D}+i\omega}.\label{eq:MM_mu_calc}
\end{equation}
Using Eqs.~(\ref{eq:T_FD_definition}), (\ref{eq:MM_D_calc}), and (\ref{eq:MM_mu_calc}) we find that the equilibrium GK formula holds with $T_{FD}(\omega)=T_{G}$, even though the system is far from equilibrium.

We now turn to the derivation of the GK relations in the Maxwell model in higher dimensions. We will discuss the three dimensional case in what follows, but the derivation applies to any dimension, and generalizes previous 1D \cite{baldassarri_2005,shokef2006} and 2D \cite{puglisi2007} results.

The equation of motion for the velocity $\mathbf{v}_{i}$ in the laboratory frame is
\begin{align}
& \mathbf{v}_{i}\left(  t+\Delta t\right) = \left\{
\begin{array}[c]{cc}
\text{\underline{value:}} & \text{\underline{probability:}}\\
\left(  1-\lambda\Delta t\right)  \mathbf{v}_{i}+\bm{\psi}_{i}\sqrt{\Delta t}
& 1-\Gamma\Delta t \\
\mathbf{v}_{i}-\frac{1+\alpha}{2}
\left( \mathbf{\hat{n}}\cdot \Delta \mathbf{v}_{ij} \right)
\mathbf{\hat{n}} & \Gamma\Delta t
\end{array}
\right.  ,\label{eq:highD_mm}
\end{align}
where the quantities in the right-hand side are taken at time $t$. As in the 1D derivation, $\lambda$ denotes the drag coefficient, the vector $\bm{\psi}_{i}$ is the random driving acceleration applied to particle $i$, and $\alpha$ is the restitution coefficient. $\mathbf{\hat{n}\equiv}\left(\mathbf{r}_{i}-\mathbf{r}_{j}\right)  /2R$ is the unit vector between the centers of the colliding particles. $R$ is the particle radius. $\Delta\mathbf{v}_{ij}\equiv\mathbf{v}_i-\mathbf{v}_j$ is the precollisional relative velocity.

As in the 1D case, the derivation of the GK relation consists of separate calculations of the mobility and autocorrelation; the GK relation is then obtained by taking the ratio of the two results. 

\subsection{Mobility}

To calculate the mobility, we add an oscillatory driving to Eq. (\ref{eq:highD_mm}):
\begin{align}
\mathbf{v}_{i}\left(  t+\Delta t\right)   &  =\nonumber\\
\bm{\xi}_{0}e^{i\omega t}\Delta t+ &  \left\{
\begin{array}[c]{cc}
\text{\underline{value:}} & \text{\underline{probability:}}\\
\left(  1-\lambda\Delta t\right)  \mathbf{v}_{i}+\bm{\psi}_{i}\sqrt{\Delta t}
& 1-\Gamma\Delta t\\
\mathbf{v}_{i}-\frac{1+\alpha}{2}\left(  \mathbf{\hat{n}}\cdot\Delta
\mathbf{v}_{ij}\right)  \mathbf{\hat{n}} & \Gamma\Delta t
\end{array}
\right.  .\label{eq:highD_EOM_mob}
\end{align}
Averaging over the stochasticity and the initial distribution
we find:
\begin{align}
\left\langle \mathbf{v}_{i}\left(  t+\Delta t\right)  \right\rangle
& = \bm{\xi}_{0}e^{i\omega t}\Delta t+\left(  1-\Gamma\Delta t\right)  \left(
1-\lambda\Delta t\right)  \left\langle \mathbf{v}_{i}\left(  t\right)
\right\rangle \nonumber\\
&  +\Gamma\Delta t\left\langle \mathbf{v}_{i}\left(  t\right)  \right\rangle
-\Gamma\Delta t\frac{1+\alpha}{2}\left\langle \left(  \mathbf{\hat{n}}%
\cdot\Delta\mathbf{v}_{ij}\right)  \mathbf{\hat{n}}\right\rangle
,\label{eq:mob_derivation1}
\end{align}
where the term involving $\bm{\psi}_{i}$ vanishes, since $\left\langle\bm{\psi}_{i}\right\rangle=0$.

\begin{figure}[ptb]
\includegraphics[width=0.8\columnwidth]{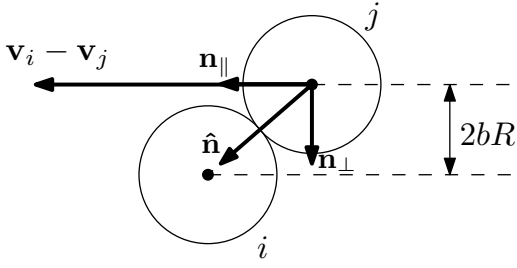}
\caption{Collision of two particles in the rest frame of particle $i$ before the collision.}\label{fig:collisiongraph}
\end{figure}

Consider the collision of two particles, $i$ and $j$. Figure \ref{fig:collisiongraph} depicts this collision, in the reference frame moving with the pre-collision velocity of particle $i$. $b$ is the dimensionless impact parameter, $0\leq b\leq1$, distributed with probability $\left(d-1\right)b^{d-2}$ in $d$ dimensions, for $d>1$ (for $d=1$, $b=0$ identically). Due to the assumption of molecular chaos, $b$\ is independent of the precollisional velocities $\mathbf{v}_{i}\left(0\right), \mathbf{v}_{i}\left(t\right), \mathbf{v}_{j}\left(t\right)$ and $\Delta \mathbf{v}_{ij}\left(t\right)$. Let us decompose$~\mathbf{\hat{n}=n}_{\shortparallel}+\mathbf{n}_{\perp}$, such that $\mathbf{n}_{\shortparallel}$ is aligned with the collision direction $\Delta\mathbf{v}_{ij}\left(t\right)$, and $\mathbf{n}_{\perp}$ perpendicular to that direction, see Fig. \ref{fig:collisiongraph}. Note that $\mathbf{n}_{\shortparallel},\mathbf{n}_{\perp}$ are not of unit length. Now define $\mathbf{\hat{k}}_{\shortparallel},\mathbf{\hat{k}}_{\perp}$ to be the unit vectors in the $\mathbf{n}_{\shortparallel},\mathbf{n}_{\perp}$ directions respectively. Then $\mathbf{n}_{\perp}=b\mathbf{\hat{k}}_{\perp}$ and $\mathbf{n}_{\shortparallel}=\sqrt{1-b^{2}}\mathbf{\hat{k}}_{\shortparallel}$. Substituting these definitions and neglecting terms of order $\left( \Delta t \right)^2$, the preceding equation reads
\begin{align}
\left\langle \mathbf{v}_{i}\left(  t+\Delta t\right)  \right\rangle
&  =\bm{\xi}_{0}e^{i\omega t}\Delta t+\left(  1-\lambda\Delta t\right)
\left\langle \mathbf{v}_{i}\left(  t\right)  \right\rangle \nonumber\\
&  -\Gamma\Delta t\frac{1+\alpha}{2}\left\langle \left(  1-b^{2}\right)
\left(  \mathbf{\hat{k}}_{\shortparallel}\cdot\Delta\mathbf{v}_{ij}\right)
\mathbf{\hat{k}}_{\shortparallel}\right\rangle \nonumber\\ 
&  -\Gamma\Delta t\frac{1+\alpha}{2}\left\langle \left(  \mathbf{n}%
_{\shortparallel}\cdot\Delta\mathbf{v}_{ij}\right)  \mathbf{n}_{\perp
}\right\rangle .\label{eq:mob_derivation2}
\end{align}
Since $\mathbf{\hat{k}}_{\shortparallel}$ and $\Delta\mathbf{v}_{ij}$ are co-aligned,
\begin{equation}
\left(\mathbf{\hat{k}}_{\shortparallel}\cdot\Delta\mathbf{v}_{ij}\right)
\mathbf{\hat{k}}_{\shortparallel}=\Delta\mathbf{v}_{ij},
\end{equation}
and because $b$ is independent of $\Delta\mathbf{v}_{ij}\left(t\right)$,
\begin{align}
&  \left\langle \left(  1-b^{2}\right)  \left(  \mathbf{\hat{k}}
_{\shortparallel}\cdot\Delta\mathbf{v}_{ij}\right)  \mathbf{\hat{k}
}_{\shortparallel}\right\rangle \nonumber\\ 
& =\left\langle \left(  1-b^{2}\right)  \Delta\mathbf{v}_{ij}\right\rangle
=\left\langle 1-b^{2}\right\rangle \left\langle \Delta\mathbf{v}
_{ij}\right\rangle .\label{eq:mob_derivation3}
\end{align}
The term $\left\langle\left(\mathbf{n}_{\shortparallel}\cdot\Delta\mathbf{v}_{ij}\right)\mathbf{n}_{\perp}\right\rangle$ 
in Eq. (\ref{eq:mob_derivation2}) vanishes (without assuming it can be decomposed into uncorrelated terms), since it is anti-symmetric in $\mathbf{n}_{\perp}$: for every $\mathbf{n}_{\perp}$, the contribution of $\left(\mathbf{n}_{\shortparallel}\cdot\Delta\mathbf{v}_{ij}\right)\mathbf{n}_{\perp}$ to the average, with $\mathbf{n}_{\perp}\rightarrow-\mathbf{n}_{\perp}$, is equally probable, and with opposite sign. Hence, using Eq. (\ref{eq:mob_derivation3}), Eq. (\ref{eq:mob_derivation2}) becomes
\begin{align}
\left\langle \mathbf{v}_{i}\left(  t+\Delta t\right)  \right\rangle
& =\bm{\xi}_{0}e^{i\omega t}dt +\left(  1-\lambda\Delta t\right)  
\left\langle \mathbf{v}_{i}\left(t\right)  \right\rangle \nonumber\\ 
& -\Gamma\Delta t\frac{1+\alpha}{2}\left\langle
1-b^{2}\right\rangle \left\langle \Delta\mathbf{v}_{ij}\right\rangle
\nonumber\\
&  =\bm{\xi}_{0}e^{i\omega t}\Delta t+ \left(  1-\lambda\Delta t\right) 
 \left\langle \mathbf{v}_{i}\left(
t\right)  \right\rangle \nonumber\\
& -\Gamma\Delta t\frac{1+\alpha}{2}\left\langle
1-b^{2}\right\rangle \left\langle \mathbf{v}_{i}\left(  t\right)
\right\rangle ,
\end{align}
since $\left\langle\mathbf{v}_{j}\left(t\right)\right\rangle=0$. Gathering the terms, we obtain the differential equation
\begin{equation}
\frac{d\left\langle\mathbf{v}_{i}\left(t\right)\right\rangle}{dt}=-\kappa\left\langle\mathbf{v}_{i}\left(t\right)\right\rangle+\bm{\xi}_{0}e^{i\omega t},\label{eq:mm_mobility_diff_equation}
\end{equation}
with $\kappa\equiv\lambda+\frac{1+\alpha}{2}\left\langle 1-b^{2}\right\rangle\Gamma$. Hence
\begin{equation}
\mu(\omega)=\frac{1}{\kappa+i\omega},\label{eq:mu_highd_result}
\end{equation}
as in the 1D case, but with a dimension-dependent effective drag coefficient $\kappa$.

\subsection{Autocorrelation}

Using Eq.~(\ref{eq:highD_mm}), we compute the change in $\left\langle\mathbf{v}_{i}\left(0\right)\mathbf{v}_{i}\left(t\right)\right\rangle$ during a short interval $\Delta t$. We multiply both sides of Eq.~(\ref{eq:highD_mm}) by $\mathbf{v}_{i}\left(0\right)$, and take the ensemble average. Noting that $\left\langle \bm{\psi}_{i}\left(t\right)\mathbf{v}_{i}\left(0\right)\right\rangle=0$, and $\mathbf{n}_{\bot}\cdot\Delta\mathbf{v}_{ij}\left(t\right)=0$, we find
\begin{align}
&  \left\langle \mathbf{v}_{i}\left(  0\right)  \mathbf{v}_{i}\left(  t+\Delta
t\right)  \right\rangle 
=\left(  1-\lambda\Delta t\right)  \left\langle \mathbf{v}_{i}\left(
0\right)  \mathbf{v}_{i}\left(  t\right)  \right\rangle \nonumber\\
&  -\Gamma\Delta t\frac{1+\alpha}{2}\left\langle \left(  \mathbf{n}%
_{\shortparallel}\cdot\Delta\mathbf{v}_{ij}\right)  \left[  \left(
\mathbf{n}_{\shortparallel}+\mathbf{n}_{\perp}\right)  \cdot\mathbf{v}%
_{i}\left(  0\right)  \right]  \right\rangle .
\end{align}
With $\mathbf{\hat{k}}_{\shortparallel}$ as defined above, this becomes
\begin{align}
&  \left\langle \mathbf{v}_{i}\left(  0\right)  \mathbf{v}_{i}\left(  t+\Delta
t\right)  \right\rangle 
=\left(  1-\lambda\Delta t\right)  \left\langle \mathbf{v}_{i}\left(
0\right)  \mathbf{v}_{i}\left(  t\right)  \right\rangle \nonumber\\
&  -\Gamma\Delta t\frac{1+\alpha}{2}\left\langle \left(  1-b^{2}\right)
\left[
\left(  \mathbf{\hat{k}}_{\shortparallel}\cdot\Delta\mathbf{v}
_{ij}\right)  \mathbf{\hat{k}}_{\shortparallel}
\right] \cdot\mathbf{v}%
_{i}\left(  0\right)  \right\rangle \nonumber\\
&  -\Gamma\Delta t\frac{1+\alpha}{2}\left\langle \left(  \mathbf{n}
_{\shortparallel}\cdot\Delta\mathbf{v}_{ij}\right)  \left[
\mathbf{n}_{\perp}\cdot\mathbf{v}_{i}\left(  0\right)  \right]  \right\rangle
.
\end{align}

As in the mobility calculation, the term $\left\langle\left(\mathbf{n}_{\shortparallel}\cdot\Delta\mathbf{v}_{ij}\right)  \left[\mathbf{n}_{\perp}\cdot\mathbf{v}_{i}\left(0\right)\right]\right\rangle$ vanishes, because it is anti-symmetric under the transformation: $\mathbf{n}_{\perp}\rightarrow-\mathbf{n}_{\perp}$. Moreover, $\left(\mathbf{\hat{k}}_{\shortparallel}\cdot\Delta\mathbf{v}_{ij}\right)\mathbf{\hat{k}}_{\shortparallel}=\Delta\mathbf{v}_{ij}$. Rearranging we find 
\begin{equation}
\frac{d\left\langle \mathbf{v}_{i}\left(  0\right) \cdot \mathbf{v}_{i}\left(t\right)\right\rangle }{dt}=-\kappa\left\langle \mathbf{v}_{i}\left(t\right) \cdot\mathbf{v}_{i}\left(0\right)\right\rangle \label{eq:highD_mm_corr}
\end{equation}
with $\kappa$ as defined following Eq. (\ref{eq:mm_mobility_diff_equation}). Hence, as in the 1D case,
\begin{equation}
D\left(\omega\right)=\frac{\langle v^{2}\rangle}{\kappa+i\omega} \label{eq:D_highd_result}
\end{equation}

Eq. (\ref{eq:D_highd_result}) and (\ref{eq:mu_highd_result}) are generalizations of Eq. (\ref{eq:MM_D_calc}) and (\ref{eq:MM_mu_calc}) respectively; Indeed, in 1D, $b=0$, and the result $\kappa_{1D}=\lambda+\frac{1}{2}\Gamma\left(  1+\alpha\right)$ is reproduced. In 2D and 3D, $\kappa_{2D}\equiv\lambda+\frac{2}{3}\Gamma\left(1+\alpha\right)$, and $\kappa_{3D}\equiv\lambda+\frac{7}{8}\Gamma\left(1+\alpha\right)$. As in the 1D case, Eq. (\ref{eq:mu_highd_result}) and (\ref{eq:D_highd_result}) show that the mobility and autocorrelation functions have the same frequency-dependence, hence the frequency-dependent GK relation holds for the Maxwell model, with $T_{FD}(\omega)=T_{G}$, in any dimension, and for both customarily studied thermostats.

\section{Mean-field granular gas}

In this section the GK relation is discussed for the MFGG model which has a velocity dependent collision rate. The zero frequency GK relation (Einstein relation) for the 3D version of this model has been studied previously. Using kinetic theory approximations, the value of $T_{FD}\left(0\right)$ was shown to be close to, but not exactly equal to $T_{G}$. For a tracer particle of the same mass and size as the rest of the particles, $T_{FD}\left(0\right)$ deviates from $T_G$ by up to 1\% for the stochastic thermostat and by up to 6\% for the Gaussian thermostat \cite{garzo}.

We now show that in the \emph{infinite} frequency limit the FD temperature is \emph{exactly} equal to the granular temperature:
\begin{equation}
\frac{T_{FD}\left(\omega\right)}{T_{G}}\underset{\omega\rightarrow\infty}{\longrightarrow}1.\label{eq:inf_freq_result}
\end{equation}
In fact, what we show is that for frequencies much higher than the highest characteristic frequency in the system, this relation holds. Eq. (\ref{eq:inf_freq_result}), together with the results of \cite{garzo}, show that the FD relation is violated. They do, however, indicate that the violation, measured as the variation in $T_{FD}\left(\omega\right)$ with frequency, is small. 

Consider first the mobility. The mobility in the Maxwell model was calculated using Eq. (\ref{eq:mm_mobility_diff_equation}). This equation must be modified for the MFGG. This is because the effective drag experienced by a particle depends on its instantaneous velocity: $\kappa=\kappa\left(\mathbf{v}\right)$. Note that, due to the mean-field (molecular chaos) assumption, the collisions of a particle do not depend on the history of the particle, and therefore $\kappa$ \emph{only depends on the current particle velocity}. The analog of Eq. (\ref{eq:mm_mobility_diff_equation}) can therefore be written as
\begin{equation}
\frac{d\left\langle\mathbf{v}_{i}\left(t\right)\right\rangle}{dt}=-\left\langle\kappa\left(\mathbf{v}_{i}\right)  \mathbf{v}_{i}\left(t\right)\right\rangle +\bm{\xi}\left(t\right),\label{eq:vel_dependent_mob}
\end{equation}
where as before the function $\bm{\xi}\left(t\right)$ is the force applied to the tracer particle. We denote the velocity distribution of the tracer particle at time $t$ by $p\left(\mathbf{v}_{i},t\right)$. Writing the averages explicitly, Eq. (\ref{eq:vel_dependent_mob}) becomes
\begin{equation}
\int\frac{\partial p\left(  \mathbf{v}_{i},t\right)  }{\partial t} \mathbf{v}_{i}d\mathbf{v}_{i}=-\int p\left(  \mathbf{v}_{i},t\right) \kappa\left(  \mathbf{v}_{i}\right)  \mathbf{v}_{i}d\mathbf{v}_{i} +\bm{\xi}\left(  t\right),
\end{equation}
and its Fourier transform reads
\begin{equation}
i\omega\int\widetilde{p}\left(  \mathbf{v}_{i},\omega\right)  \left[1+\frac{\kappa\left(  \mathbf{v}_{i}\right)  }{i\omega}\right]  \mathbf{v}_{i}d\mathbf{v}_{i}=\widetilde{\bm{\xi}}\left(  \omega\right) ,
\end{equation}
where $\widetilde{p}\left(\mathbf{v}_{i},\omega\right),\widetilde{\bm{\xi}}\left(\omega\right)$ are the Fourier transforms of $p\left(\mathbf{v}_{i},t\right),\bm{\xi}\left(t\right)$ respectively. Rearranging, and noting that $\left\langle \mathbf{\tilde{v}}_{i}\left(\omega\right)\right\rangle$ and $\widetilde{\bm{\xi}}\left(\omega\right)$ are vectors in the same direction, so that their quotient is well-defined, we find that
\begin{equation}
\mu(\omega)\equiv\frac{\left\langle \mathbf{\tilde{v}}_{i}\left(
\omega\right)  \right\rangle }{\widetilde{\bm{\xi}}\left(  \omega\right)
}=\frac{\int\widetilde{p}\left(  \mathbf{v}_{i},\omega\right)  \mathbf{v}
_{i}d\mathbf{v}_{i}}{\widetilde{\bm{\xi}}\left(  \omega\right)  }=\frac
{1}{i\omega}+O\left(  \frac{1}{\omega^{2}}\right)  ,\label{eq:mu_w_order_w}
\end{equation}
where $O\left(\frac{1}{\omega^{2}}\right)$ denotes a function of order $\left(1/\omega\right)^{2}$, for $1/\omega\rightarrow0$. 

To calculate the fluctuation $D\left(\omega\right)$, we denote: $C\left(t\right)~\equiv~\langle\mathbf{v}(0)\mathbf{v}(t)\rangle$, and the Fourier-Laplace transform of a function by $FL\left[  ..\right]$. Then:
\begin{align}
FL\left[C\left(t\right)\right] & =\int_{0}^{\infty}C\left(t\right)e^{-i\omega t}dt\nonumber\\
&  =\frac{-1}{i\omega}\left[  \left.  C\left(  t\right)  e^{-i\omega t}\right\vert_{t=0}^{\infty}-\int_{0}^{\infty}\frac{dC\left(t\right)}{dt}e^{-i\omega t}dt\right] \nonumber\\
&  =\frac{1}{i\omega}\left\{C\left(0\right)+FL\left[\frac{dC\left(t\right)}{dt}\right]\right\}.
\end{align}
This recursion relation \cite{footnote} is satisfied by the infinite series
\begin{equation}
D\left(  \omega\right)  =\frac{1}{d}FL\left[  C\left(  t\right)  \right] =\frac{1}{d}\sum_{n=0}^{\infty}\frac{1}{\left(  i\omega\right)  ^{n+1}}\left. \frac{d^{n}C\left(  t\right)  }{dt^{n}}\right\vert _{t=0^{+}}.
\end{equation}
If all the derivatives of $C\left(t\right)$ are well-behaved, we have
\begin{equation}
D\left(  \omega\right)  =\frac{1}{d}\frac{1}{i\omega}C\left(  0\right)+O\left(  \frac{1}{\omega^{2}}\right). \label{eq:d_w_order_w}
\end{equation}
Using Eq. (\ref{eq:mu_w_order_w}) and (\ref{eq:d_w_order_w}), the resulting FD temperature is
\begin{equation}
T_{FD}\left(\omega\right) = \frac{D\left(\omega\right)}{\mu\left(\omega\right)}\overset{\omega\rightarrow\infty}{\rightarrow}\frac{1}{d}C\left(  0\right)=\frac{1}{d}\left\langle \mathbf{v}^{2}\right\rangle=T_{G},
\end{equation}
which is what we wanted to show; c.f. Eq. (\ref{eq:inf_freq_result}).

This result can be interpreted as follows: the system possesses two time scales, the mean time between collisions, and the thermostat time scale, $\lambda^{-1}$.\ For frequencies higher than the inverse of these two time scales, $D\left(\omega\right)$ and $\mu\left(\omega\right)$ have the same frequency dependence, and the FD relation holds, with $D\left(\omega\right)/T_{G}\simeq\mu\left(\omega\right)\simeq\left(i\omega\right)^{-1}$.

\begin{figure}[ptb]
\includegraphics[width=\columnwidth]{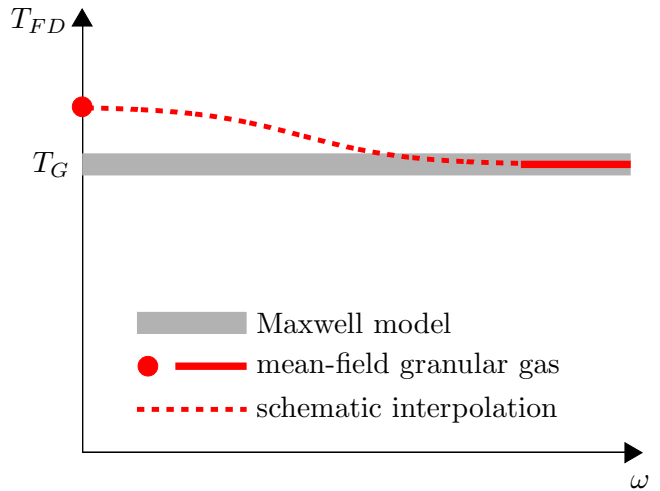}
\caption[Fluctuation dissipation in the Maxwell model, and in a varying collision rate model. ]{(Color online) FD temperature in the Maxwell model, and in the mean-field granular gas model. In the former, the temperature obtained from the Green-Kubo relations is frequency independent and equal to the granular temperature. In the latter, the temperature obtained depends on the measurement frequency. The schematic curve interpolates between the high and low $\omega$ limits. The difference between the two curves is not to scale.}\label{maxmod_garzo_violation}
\end{figure}

\section{Discussion}

The results for the two models can be summarized as follows, see Fig. \ref{maxmod_garzo_violation}. In the Maxwell model the FD relation holds, with a single FD temperature for all frequencies. In the more physically realistic MFGG, the FD relation is violated: whereas \cite{garzo} gives $T_{FD}\left(0\right)\neq T_{G}$, we have found $\lim_{\omega\rightarrow\infty}T_{FD}\left(\omega\right)=T_{G}$, demonstrating that $T_{FD}$ is not independent of $\omega$. The difference in $T_{FD}$ between the limits of high and low frequency is smaller than several percent. We would thus expect that the variation in $T_{FD}\left(\omega\right)$ is small for all frequencies. This variation is expected to be the largest for small $\alpha$ and could perhaps be detected in simulations. Molecular dynamics simulations are hard to conduct for very inelastic particles ($\alpha \approx 0$), since the grains tend to cluster \cite{Goldhirsch,mcnamara_young,luding}. In direct simulation Monte Carlo techniques, on the other hand, the mean-field property is explicitly imposed, and the variation of $T_{FD}$ with frequency may be easier to detect. It would furthermore be interesting to explore whether analytic progress may be achieved in studying the high frequency limit for dense systems where correlations give rise to significant FD violations \cite{puglisi2007}. Last, we note that the technique we used to calculate the infinite frequency limit for the MFGG model applies to a more general class of models, such as models in which the restitution coefficient depends on the particles' relative velocity \cite{luding_1996,bizon,mcnamara_falcon}.

\acknowledgments

We thank Erez Braun, Naama Brenner, Robert Dorfman, Yariv Kafri, Andrea Liu, and Tom Lubensky for helpful discussions. This work was supported by Grant No. 660/05 of the Israel Science Foundation, the Fund for the Promotion of Research at the Technion, and NSF MRSEC program under Grant No. DMR 05-20020.


\begin{thebibliography}{99}
\bibitem {footnote} Note that this relation may be equally well obtained by performing a Fourier-Laplace 
transform on both sides of a Taylor-MacLauren expansion of $C$: $C(t) = \sum \frac{1}{n!} \frac{d^{n} C}{dt^{n}}|_{0}t^{n}$.
\end{thebibliography}
\end{document}